# Data Visualization on Shared Usage Multi-Screen Environment

Ph.D. Yuriy A. Chashkov

MNTC "ATLANT" (Moscow, Russia)
E-mail: chashkov@mail.ru

**Abstract.** The modern multimedia technologies based on the whole palette of hardware and software facilities of real-time high-speed information processing, in a combination with effective facilities of the remote access to information resources, allow us to visualize diverse types of information. Data visualization facilities – is the face of the Automated Control System on whom often judge about their efficiency. They take a special place, providing visualization of the diverse information necessary for decision-making by a final control link - the person allocated by certain powers.

At creation of Shared Usage Multi-Screen Environment (SUME) one of the basic technical problem - data visualizing process management. However there are cases, when document loading process or displaying service information (for example, windows headers, service panels, etc.) is extremely undesirable. Decision of this problem - development special client-server applications which allows to operate with videoserver through LAN or communication port.

If SUME installed in the conference hall which used in commercial objectives for carrying out various business presentations, it is possible to use videoserver under the control of any Windows platform. But, if SUME used as Crisis Center at the territory with restricted access, where there are special requirements on software certification and information security, - as videoserver operational system it is necessary to use only certificated Operational Systems. In that case most rational and universal decision - using Java technology.

Unfortunately, here again there are lacks main of which - impossibility to get direct access to DCOM objects, so also to any files created with Microsoft Office and the majority of other applications. To not go deep into the problem of data storage formats studying it is necessary to use special bridges.

Bridges it's pure Java implementation of DCOM enables programming with COM objects from any platform which supports the Java Developer Kit (JDK). Visual Studio developers also benefit from the ability to access any Java object as if it were implemented as a COM object. For today it is known such three projects: Intrinsys J-Integra [1], EZ-JCom [2], IBM Interface Tool for Java [3].

Bridges allows developers to work in their respective environments, while having the benefit of accessing COM components from Java, and Java components from COM (fig.1). The ultimate benefit is that developers can save time by leveraging existing software code, delivering best-in-class software solutions, and avoiding the need to learn new programming environments.

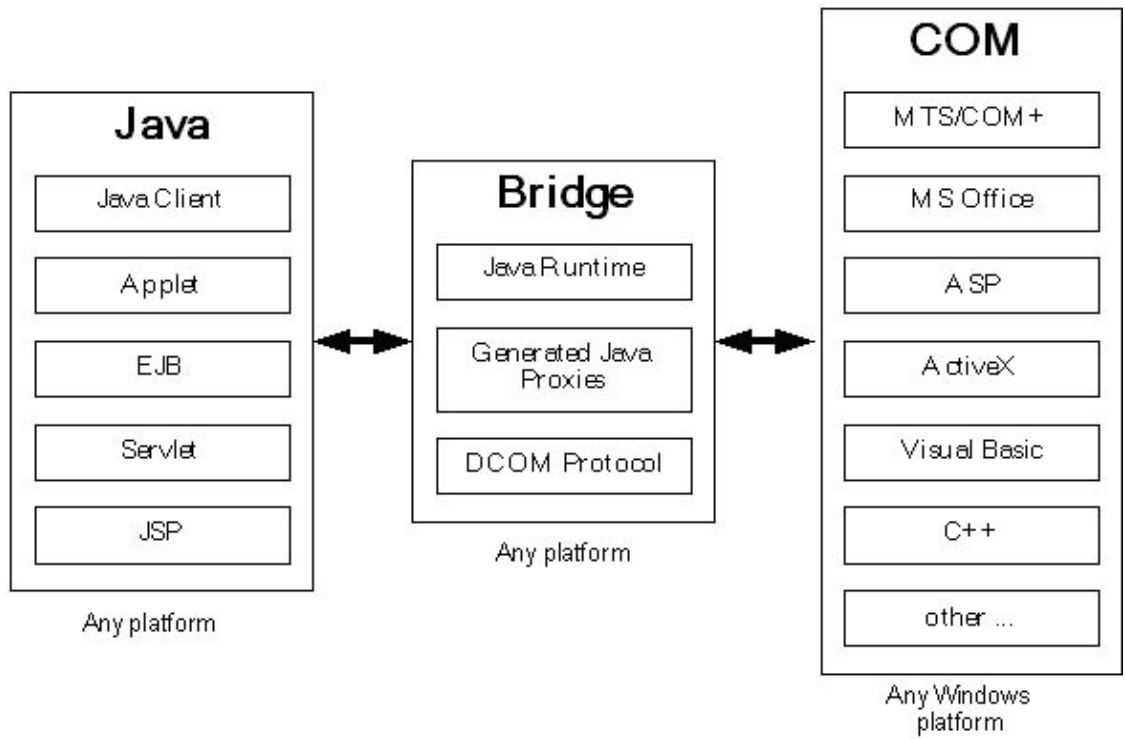

Fig.1.

Bridge gives Java access to COM components by generating Java proxy classes from a COM component's type library (TLB). These proxy classes expose COM properties, methods, and events as their Java equivalents. They also supports early binding access to Java objects by generating a TLB file from a Java object's class files. These TLB files expose Java properties, methods, and events as their COM equivalents.

Bridges will give the power to:

- Write clients for EJBs (Enterprise Java Beans) in any COM-supported language such as Visual Basic or C++;

- Access COM components from any type of Java client, including Applets, EJBs, Servlets, JSPs, and standalone applications;

- Cleanly integrate two of the leading technologies in one environment without rewriting any code;

- Maximize reuse of existing Java and COM components;

- Leverage the existing skills base of your workforce. For example, move to Visual Basic for your client side technology while continuing to use J2EE server.

Because its bridging capabilities are infinitely flexible, bridges can be used from any COM client environment and can access virtually every kind of COM object. Some uses of bridges

include:

- Providing COM/Java ARs to products;

- Accessing Enterprise JavaBeans (EJBs) from Active Server Pages (ASP);

- Populating Microsoft Office documents from Java code;

- Creating Visual Basic clients for EJBs;

- Accessing Microsoft Exchange from JavaServer Pages;

- Generating Microsoft Word documents from Java code.

A pure Java implementation means that bridges runs on any JVM. With its mapping of COM types to Java native types there is no need to learn a new type system or write type-converting code. With a pure Java implementation of DCOM, developers are no longer bound to a single platform. An elegant mapping of COM types to Java types means there is no need to learn a new type system or write type-converting code. For example, all COM interfaces and types are presented to Java developers as Java interfaces and classes. Methods in the COM interface become member functions of the corresponding Java class, and COM properties are mapped to their equivalent getter/setter methods on the Java side. Hence all the complexities of COM and DCOM are hidden from the Java developer. Exceptions raised in Visual Basic can be caught in Java code using the standard try/catch syntax.

For example let's show code fragment which open and start PowerPoint presentation:

```
import powerpoint.*;         // import generated PowerPoint proxy
import java.io.*;
import java.util.*;

public class SlideShowIBM {

        public SlideShowIBM (String fileName,
            int XPosition, int YPosition,
            int width, int height) {

          Application app = null;
          Presentations presentations = null;
          Presentation presentation = null;
          SlideShowSettings slideShowSettings = null;
          Slide slide = null;
          SlideShowWindow showWindow = null;
          SlideShowView showView = null;

            try {
              // Initialize the Java2Com Environment
              com.ibm.bridge2java.OleEnvironment.Initialize();
```

```
            app = new Application ();
            // set application minimized
            app.set_WindowState(2);
            app.set_Visible (1);   // set application visible
            presentations = app.get_Presentations ();
            presentations.Open (fileName, 0, 0, 0);
            presentation =
              presentations.Item (new Integer(1));
            slideShowSettings =
              presentation.get_SlideShowSettings();
            slide =
              presentation.get_Slides().Add(1,
                PpSlideLayout.ppLayoutTitle);

            showWindow = slideShowSettings.Run();
            showWindow.set_Width ((float)width);
            showWindow.set_Height ((float)height);
            showWindow.set_Left ((float)XPosition);
            showWindow.set_Top ((float)YPosition);

            showView = showWindow.get_View ();
            Thread.sleep (1000000);
            app.Quit();
        } catch (com.ibm.bridge2java.ComException e) {
           System.out.println ("COM Exception");
           System.out.println (e.getMessage());
           System.out.println (e);
        } catch (Exception e) {
           System.out.println ("message: " + e.getMessage());
        } finally {
           app = null;
           com.ibm.bridge2java.OleEnvironment.UnInitialize();
        }
     }
 // --------------------------------------------
        public static void main (String [] args) {
         new ShowPP ("d:\\presentation.ppt", 300, 200, 50, 100);
        }
 }
```

## REFERENCES

1. http://www.intrinsyc.com

2. http://www.ezjcom.com

3. http://www.alphaworks.ibm.com/tech/bridge2java